\documentclass[12pt]{article}

\usepackage{graphicx}
\begin{document}

\begin{center}
{\bf Nonlinear electrodynamics without singularities}\\
\vspace{5mm} S. I. Kruglov
\footnote{E-mail: serguei.krouglov@utoronto.ca}
\vspace{3mm}

\textit{Department of Physics, University of Toronto, \\60 St. Georges St.,
Toronto, ON M5S 1A7, Canada\\
Canadian Quantum Research Center, \\
204-3002 32 Ave., Vernon, BC V1T 2L7, Canada} \\
\vspace{5mm}
\end{center}

\begin{abstract}

Nonlinear electrodynamics with two parameters is studied. It is shown that singularities of point-like electric charges are absent and the electromagnetic energy is finite. Corrections to Coulomb's law are found. The finite static electric field energy of a point-like charge is calculated.  We demonstrate that the electron mass may have the electromagnetic nature. It was shown that principles of causality and unitarity take place.

\end{abstract}

First nonlinear electrodynamics (NED) was proposed by Born and Infeld \cite{Born}.  That NED removes a singularity of point-like charges and electric self-energy is finite. Quantum electrodynamics (QED) takes into account quantum corrections and it becomes NED \cite{Heisenberg,Schwinger,Adler}. There are many models of NED: Born--Infeld-like \cite{Kr.,Kr.1}, logarithmic \cite{Soleng}, arcsin-electrodynamics \cite{Kr.2}, exponential electrodynamics \cite{Hendi,Kr.3} and others. They are used in cosmology to explain Universe inflation, the existence of dark energy and the thermodynamic behaviour of  black holes. Each model gives different corrections to Coulomb's law, values of self-energy and birefringence effects. In strong fields the usage of different models  leads to different  behaviour that can, in principle, be verified in experiments to choose a valuable model. We will show that our NED has attractive
features such as: singularities of point-like electric charges are absent; the electrostatic energy is finite; there are corrections to Coulomb's
law; the electron mass may have the electromagnetic nature; principles of causality and unitarity take place.
There exist other theories of electrodynamics like the Bopp--Podolsky (BP) electrodynamics \cite{Bopp,Podolsky} and nonlocal electrodynamics. But these theories have another properties compared to NED. Thus, BP electrodynamics is linear and a higher derivative field theory. The reasons to study BP theory are to improve renormalization properties and to remove ultraviolet divergences. However, this model suffers such a difficulty as the presence of ghosts. This leads to the violation of unitarity. Some aspects of BP electrodynamics were considered in Ref. \cite{Kr.4}. Nonlocal electrodynamics uses the integro-differential operators and a general fractional calculus \cite{Tarasov,Efimov}. But in these models there is non-locality in space and time. In addition, there exists uncertainty in the chose of kernels. In this letter we study NED with two parameters which is converted into Maxwell's electrodynamics for weak electromagnetic fields.

We will study the model of NED with the Lagrangian
\begin{equation}
{\cal L}=-\frac{{\cal F}}{1+\epsilon(2\epsilon\beta {\cal F})^\gamma},
\label{1}
\end{equation}
where ${\cal F}=F^{\mu\nu}F_{\mu\nu}/4=(B^2-E^2)/2$, and $E$ and $B$ are the electric and magnetic fields, correspondingly, and $\epsilon=\pm 1$. We imply that $\beta>0$, $\gamma>0$. For $B>E$ we use $\epsilon=1$ \cite{Kruglov} and for $B<E$ we will put $\epsilon=-1$ to have the real Lagrangian. Below, it will be shown that for $\epsilon=-1$ the electric field at the origin is finite. At the limit $\beta\rightarrow 0$ Lagrangian (1) is converted into the Maxwell's Lagrangian. At $\gamma=1$, $\epsilon=1$ Eq. (1) becomes the  Lagrangian of rational NED \cite{Kruglov1}. The model of NED (1) was explored in \cite{Kruglov2,Kruglov3,Kruglov4,Habib,Habib1} for some values of $\gamma$-parameter.

The Euler--Lagrange equation (for the general theory of relativity case) follows from Eq. (1)
\begin{equation}
\nabla_\mu({\cal L}_{\cal F}F^{\mu\nu})=0,
\label{2}
\end{equation}
where
\begin{equation}
{\cal L}_{\cal F}=\frac{\partial {\cal L}}{\partial {\cal F}}=-\frac{1-\epsilon(\gamma-1)(2\epsilon\beta {\cal F})^\gamma}{(1+\epsilon(2\epsilon\beta {\cal F})^\gamma)^2}.
\label{3}
\end{equation}
With spherical symmetry Eq. (2) for the electric field ($\epsilon=-1$, ${\cal F}=-E^2(r)/2$) becomes
\begin{equation}
\frac{1}{r}\frac{d(r^2E(r){\cal L}_{\cal F})}{dr}=0.
\label{4}
\end{equation}
Making use of Eq. (3) and solving Eq. (4) we obtain
\begin{equation}
\frac{E(r)(1+(\gamma-1)(\beta E^2(r))^\gamma)}{(1-(\beta E^2(r))^\gamma)^2}=\frac{Q}{r^2},
\label{5}
\end{equation}
where $Q$ is the integration constant which we identify with the electric charge. As $\beta\rightarrow 0$ in Eq. (5) the electric field reduces to Coulomb’s field $E_C(r)=Q/r^2$.
Defining unitless variables
\begin{equation}
x=\frac{r}{\beta^{1/4}\sqrt{Q}}, ~~~~~y=\sqrt{\beta}E,
\label{6}
\end{equation}
we represent Eq. (5) as follows:
\begin{equation}
\frac{y(1+(\gamma-1)y^{2\gamma})}{(1-y^{2\gamma})^2}=\frac{1}{x^2}.
\label{7}
\end{equation}
Then, from Eq. (7), one has for small $x$ (and small $r$)
\begin{equation}
y=1-\frac{x}{2\sqrt{\gamma}}+{\cal O}(x^2).
\label{8}
\end{equation}
There are two solutions to Eq. 7 with $\pm x$. But the physical solution is with sign minus of the second term in Eq. 8.
Making use of Eqs. (6) and (8) we obtain
\begin{equation}
E(r)=\frac{1}{\sqrt{\beta}}-\frac{r}{2\sqrt{Q\gamma}\beta^{3/4}}+{\cal O}(r^2)~~~~~~r\rightarrow 0.
\label{9}
\end{equation}
Thus, one has the finite value of the electric field at the origin $E(0)=1/\sqrt{\beta}$ which is the maximum of the electric field.
The plot of $y$ versus $x$ is depicted in Fig. 1 for $\gamma=0.25,0.5,0.75$.
\begin{figure}[h]
\includegraphics{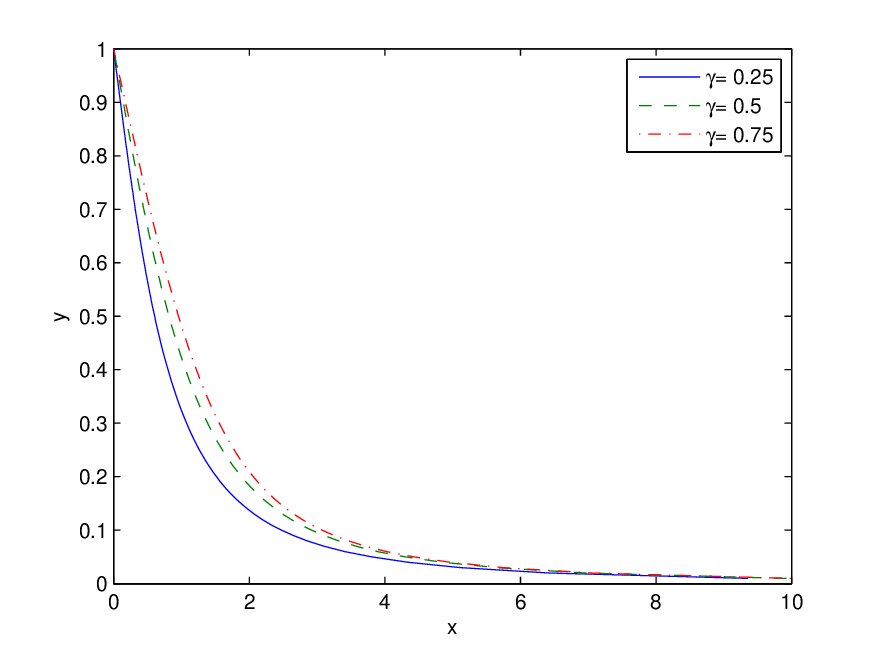}
\caption{\label{fig.1} The function $y$ vs. $x$ at $\gamma=0.25,0.5,0.75$.}
\end{figure}
As $r\rightarrow \infty$ we obtain from Eq. (5)
\begin{equation}
E(r)=\frac{Q}{r^2}+{\cal O}(r^{-4}).
\label{10}
\end{equation}
Thus, we have corrections to Coulomb's law in the order of ${\cal O}(r^{-4})$. We plotted in Fig. 1 only the physical branch when the electric field approaches to zero as $x\rightarrow\infty$  ($r\rightarrow\infty)$. The second branch with the field growing with the distance from its source is not physical. Figure 1 shows that the electric field is finite, $y=1$, at $r=0$ and goes to zero as $r\rightarrow\infty$.

The stress-energy tensor corresponding to Lagrangian (1) is given by
\begin{equation}
T_\mu^{~\nu}=-{\cal L}_{\cal F}F_{\mu\lambda}F^{\nu\lambda}-{\cal L}\delta_\mu^{~\nu}.
\label{11}
\end{equation}
Making use of Eq. (11) we obtain the electric energy density
\begin{equation}
\rho=T_0^{0}=-E^2{\cal L}_{\cal F}-{\cal L}=\frac{E^2[1+(2\gamma-1)(\beta E^2)^\gamma]}{2[1-(\beta E^2)^\gamma]^2}.
\label{12}
\end{equation}
With the help of dimensionless variables (6) the electric energy density becomes
\begin{equation}
\rho=\frac{y^2[1+(2\gamma-1)y^{2\gamma}]}{2\beta[1-y^{2\gamma}]^2}.
\label{13}
\end{equation}
The total electric energy is given by
\[
{\cal E}=\int_0^\infty \rho(r)r^2dr
\]
\begin{equation}
=\frac{Q^{3/2}}{\beta^{1/4}}\int_0^1\frac{[(2\gamma-1)y^{2\gamma}+1][2(\gamma-0.5)
\left((\gamma+2)y^{2\gamma}+(\gamma-0.5)y^{4\gamma}\right)+1]}{4\sqrt{y}[(\gamma-1)y^{2\gamma}+1]^{5/2}}dy,
\label{14}
\end{equation}
where we  have used Eq. (7). Numerical calculations of dimensionless variable $\bar{{\cal E}}\equiv {\cal E}\beta^{1/4}/Q^{3/2}$ are presented in Table 1.
\begin{table}[ht]
\caption{Approximate values of $\bar{{\cal E}}\equiv {\cal E}\beta^{1/4}/Q^{3/2}$}
\centering
\begin{tabular}{c c c c c c c c c c c}\\[1ex]
\hline
$\gamma$ & 0.1 & 0.2 & 0.3  & 0.4 & 0.5 & 0.6 & 0.7 & 0.8 & 0.9 & 1\\[0.5ex]
\hline
$\bar{{\cal E}}$ & 0.5757 &0.7376 & 0.8323 & 0.8962 & 0.9428 & 0.9786 & 1.0072 & 1.0305 & 1.0500 & 1.0667\\[0.5ex]
\hline
\end{tabular}
\end{table}
Thus, the electrostatic energy of charged objects is finite in our model of NED. One can explore the Abraham and Lorentz idea that the electron mass ($m_e={\cal E}\approx 0.51$ MeV) is the electromagnetic energy \cite{Born,Rohrlich,Spohn} to obtain the parameter $\beta^{1/4}$ in fm for some values of $\gamma$. Dirac also considered the possibility that the electron may be classical charged object \cite{Dirac}.

When causality and unitarity principles take place the NED models are viable \cite{Shabad}. The causality principle tells that a group velocity of elementary excitations over a background field is not greater than the speed of light in vacuum.  The propagator residue should be positive that is the unitarity principle. These two principles, in our notations, hold if \cite{Shabad}
\begin{equation}
 {\cal L}_{\cal F}\leq 0,~~~~{\cal L}_{{\cal F}{\cal F}}\geq 0,~~~~{\cal L}_{\cal F}+2{\cal F} {\cal L}_{{\cal F}{\cal F}}\leq 0.
\label{15}
\end{equation}
From Eqs. (1) and (3) we find ($\epsilon=-1$, $2{\cal F}=-E^2$)
\begin{equation}
{\cal L}_{\cal F}=-\frac{1+(\gamma-1)(\beta E^2)^\gamma}{(1-(\beta E^2)^\gamma)^2},~~~~
 {\cal L}_{{\cal F}{\cal F}}=\frac{2\gamma (\beta E^2)^\gamma\left[(\gamma-1)(\beta E^2)^\gamma+\gamma+1\right]}{E^2\left(1-(\beta E^2)^\gamma\right)^3}.
\label{16}
\end{equation}
Because the maximum of the electric field is $E(0)=1/\sqrt{\beta}$ we have $(\beta E^2(r))^\gamma<1$. As a result, all three inequalities in Eq. (15) are satisfied for $\gamma\leq 1$. Thus, the principles of causality and unitarity occur for any electric fields if $\gamma\leq 1$.

A model of NED, without a singularity of the electric field in the centre of charges and electric self-energy is finite, has been considered.
The principles of causality and unitarity were studied. It was shown that causality and unitarity hold.
We obtained corrections to Coulomb's law that are in the order of ${\cal O}(r^{-4})$. It is worth noting that there  exists a class of models, including quadratically truncated Euler--Heisenberg Lagrangian of QED, with finite field energy of a point charge, although the field is infinite \cite{Gitman,Shabad1} (see also \cite{Kruglov5}).

\end{document}